\documentclass[galaxies,article,submit,moreauthors,pdftex,10pt,a4paper]{mdpi} 
\preto{\abstractkeywords}{\nolinenumbers}%Eliminates line numbers

\firstpage{1} 
\articlenumber{x}
\doinum{10.3390/------}
\pubvolume{xx}
\pubyear{2017}
\copyrightyear{2017}
\history{Received: date; Accepted: date; Published: date}
 \theoremstyle{mdpi}
 \newcounter{thm}
 \newcounter{ex}
 \newcounter{re}
 \theoremstyle{mdpidefinition}

\usepackage{amssymb}
\Title{
Multiwavelength Observations of Relativistic Jets from General Relativistic Magnetohydrodynamic Simulations}

\Author{Richard Anantua $^{1,2
}$*, Roger Blandford $^{2
}
$ and Alexander Tchekhovskoy $^{1,3}$}
\AuthorNames{Richard Anantua, Roger Blandford and Alexander Tchekhovskoy}

\address{%
$^{1}$ \quad Astronomy Department, University of California, Berkeley, 601 Campbell Hall, Berkeley, CA 94720, USA; ranantua@berkeley.edu\\
$^{2}$ \quad Kavli Institute for Particle Astrophysics and Cosmology, Stanford University\, P.O. Box 20450, MS 29, Stanford, CA 94309, USA; rdb3@stanford.edu\\
$^{3}$ \quad Department of Physics and Astronomy and CIERA, Northwestern University, 2145 Sheridan Road, Evanston, IL 60208, USA; atchekho@northwestern.edu
}

\corres{Correspondence: ranantua@berkeley.edu
}

\abstract{
This 
work 
summarizes a program intended to unify three burgeoning branches of the high-energy astrophysics of relativistic jets:
general relativistic magnetohydrodynamic (GRMHD) simulations of ever-increasing dynamical range, the microphysical theory of particle acceleration under relativistic conditions and multiwavelength observations resolving ever-decreasing spatiotemporal scales. The process, which involves converting simulation output into time series of images and polarization maps that can be directly compared to observations, is performed by: 1.) self-consistently prescribing models for emission, absorption and particle acceleration; and 2.) performing time-dependent polarized radiative transfer. M87 serves as an exemplary prototype for this investigation due to its 
prominent and well-studied jet and the imminent prospect of learning much more from Event Horizon Telescope (EHT) observations this year. Synthetic observations 
can 
be directly compared with real observations for observational signatures such as %black hole photon rings, MHD disk and/or 
jet instabilities, 
collimation, relativistic beaming and 
polarization. The simplest models described adopt the standard equipartition hypothesis;
 other models calculate emission by relating it to
current density or shear. These models are intended for application %to the 
to the radio jet instead of the higher frequency emission, the disk  and the wind, which will be subjects of future investigations.}


\begin{document}

%%%%%%%%%%%%%%%%%%%%%%%%%%%%%%%%%%%%%%%%%%
%% Sections that are not mandatory are listed as such. The section titles given are for Articles. Review papers and other article types have a more flexible structure. 

%% Only for the journal Gels: Please place the Experimental Section after the Conclusions

%%%%%%%%%%%%%%%%%%%%%%%%%%%%%%%%%%%%%%%%%%
% \setcounter{section}{-1} %% Remove this when starting to work on the template.
% \section{How to Use this Template}

% The template details the sections that can be used in a manuscript. Sections that are not mandatory are listed as such. The section titles given are for Articles. Review papers and other article types have a more flexible structure. For any questions, please contact the editorial office of the journal or support@mdpi.com. For LaTeX related questions please contact Janine Daum at latex-support@mdpi.com.

\section{Introduction}

% The introduction should briefly place the study in a broad context and highlight why it is important. It should define the purpose of the work and its significance. The current state of the research field should be reviewed carefully and key publications should be cited. Please highlight controversial and diverging hypotheses when necessary. Finally, briefly mention the main aim of the work and highlight the main conclusions. As far as possible, please keep the introduction comprehensible to scientists outside your particular field of research. Citing a journal paper \cite{ref-journal}. And now citing a book reference \cite{ref-book}.

% \section{Introduction}

% \subsection{Jet-Accretion Disk-Black Hole (JAB) Systems}
% Disks of accreting matter may be found orbiting gravitating bodies from planetary to stellar to galactic scales throughout the observable universe. These accretion disks-- containing baryonic matter in the form of particles, dust, asteroids, planets and stars--  play an essential role in star formation, structure formation and angular momentum transport in galaxies \cite{Shakura1973}.

Relativistic jets are powerful, collimated outflows launched from compact objects typically surrounded by accretion disks in black hole X-ray binaries, gamma ray bursts or active galactic nuclei (AGN) throughout the observable universe. 
%relativistic jets
% %also ubiquitous. 
% are among the greatest sources of continuous energy output in the observable universe. 
Of particular interest in observational astronomy are relativistic jets from AGN, which 
%radiate 
% supply 
are associated with the greatest total energy output among known astrophysical sources. Ever since Heber Curtis observed “a thin line of matter” flowing from the center of M87 in 1918 \cite{Curtis1918}, AGN jet observations have proliferated, as seen in the $\mathit{Fermi}$ Gamma-ray Space Telescope catalog and Caltech's corresponding Owens Valley Radio Observatory (OVRO) 40-mm telescope radio survey of 
approximately 
%%%1200
1,200 blazar sources. 

% \subsection{Observations}

Interest in 
%JAB systems 
jets has been spurred by recent discoveries relating to their central engines-- black holes-- 
% which have been probed in novel ways following the detection of gravitational radiation from neutron star and stellar black hole mergers.
including the monumental observational confirmation of Einstein's prediction of gravitational waves from neutron star and stellar black hole mergers. Pending discoveries relating to the nature of 
%JAB system 
jet emission close to the central engine are rapidly garnering a similar level of interest \cite{Blandford2017}. Some planned observations spurring theoretical progress are summarized as follows.

% \subsubsection{Sgr A*}
The black hole in our Galactic Center, Sgr A*, subtends an angular width at Earth of 5.3 $\mu$as \cite{Christian2015}. A network of intercontinental radio  baselines anchored by the Atacama Large mm- and submm-Array (ALMA), is imaging Sgr A* at 230 GHz. The EHT is also expected to provide our first direct observations of the black hole shadow of an extragalactic source such as the misaligned BL Lac blazar M87 \cite{Lu2014}. 

\begin{figure}\nonumber
\begin{align}
 & \hspace{2cm}\includegraphics[height=120pt,trim = 6mm 1mm 0mm 1mm]{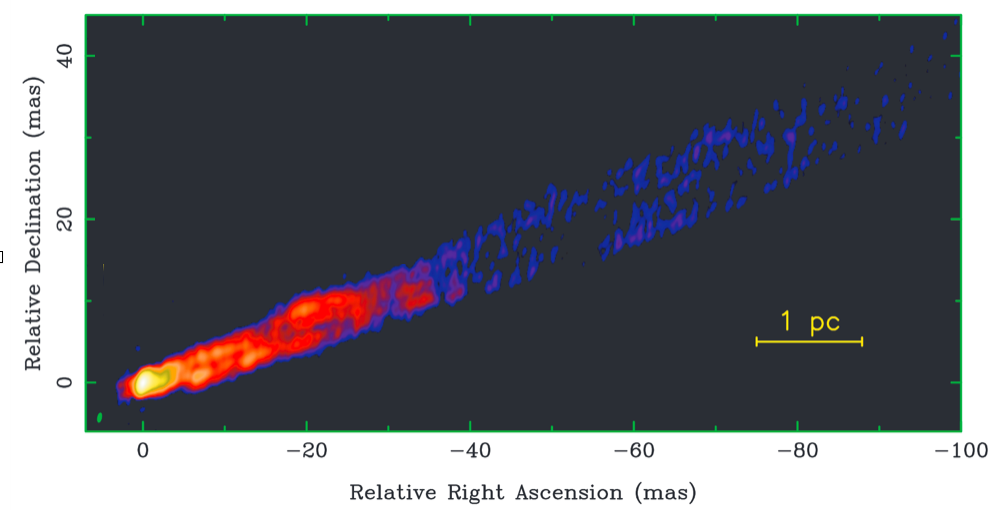}
  &  & 
\end{align}
\caption{VLBA 15 GHz (2 cm) image of M87. The swirling jet substructure 
(twisting about the jet axis)
may be indicative of magnetic Kelvin-Helmholtz instabilities. Image adapted with permission of Dan Homan, Yuri Kovalev, Matt Lister and Ken Kellermann 
%of the MOJAVE collaboration
.
}\label{SwirlingJet}
\end{figure}

\begin{figure}
\begin{center}
% Inclusion d'une image: ce qui est entre crochets est un argument "optionnel", alors que l'argument obligatoire est entre accolades
\includegraphics[width=75mm,height=50mm]{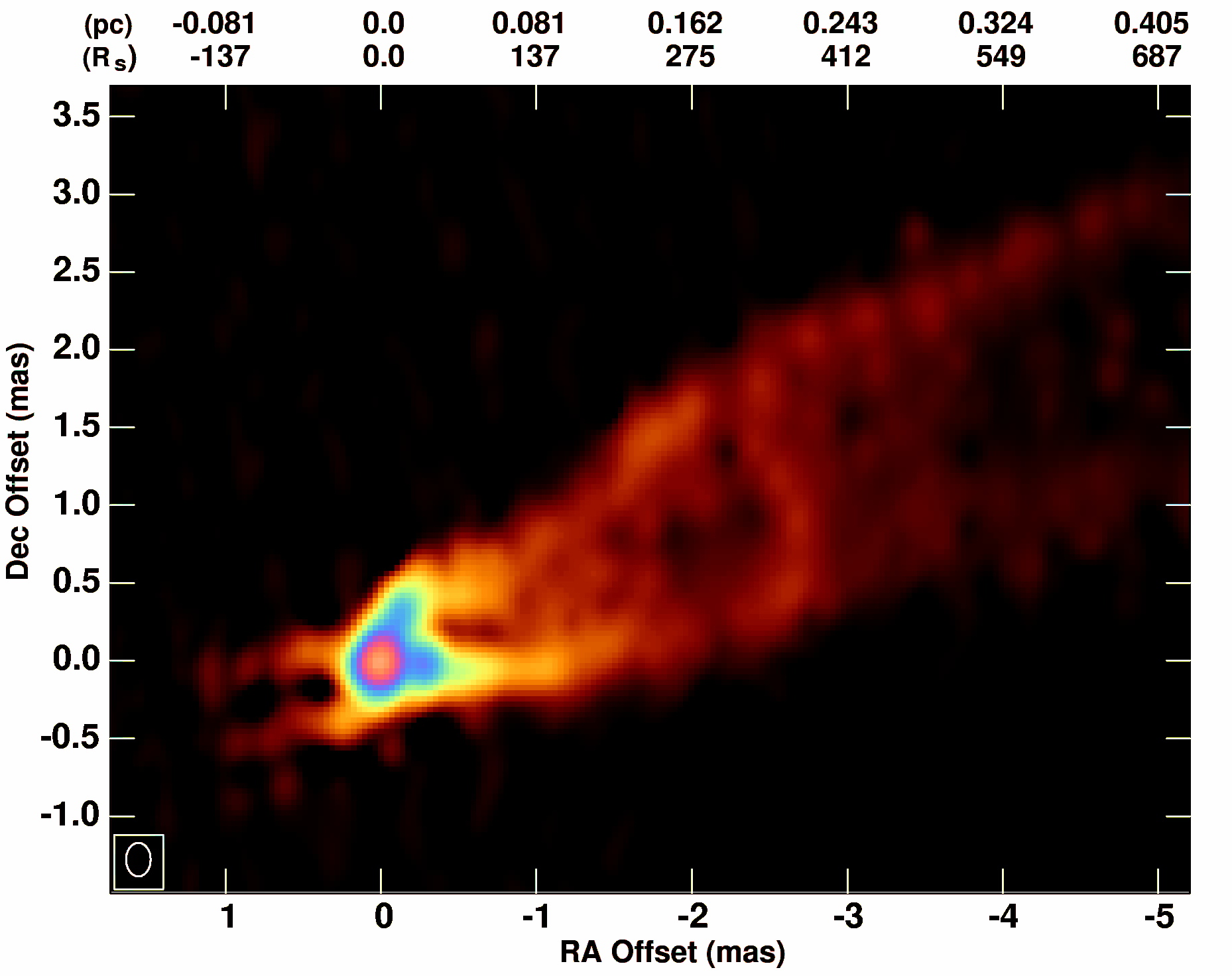}
\caption{M87 43 GHz radio image of the inner region, including dominant and counterpropagating jets. Note 1 pc $\approx 12.5$ mas. Image courtesy of Craig Walker.}\label{fig:CWalkerM87}
\end{center}
\end{figure}

% \subsubsection{M87}
M87 is a giant elliptical galaxy in the Virgo Cluster 54 million lightyears (17 Mpc) away also possessing a large (3.9 $\mu$as) \cite{Christian2015} central black hole. M87 observations by the National Radio Astronomy Organization Very Long Baseline Array (NRAO VLBA) at 15 GHz \cite{Kovalev2007} reveal substructure indicative of an ordered, helical magnetic field as shown in Figure \ref{SwirlingJet}.  %%%VLA
VLBA observations at 43 GHz have shown limb brightening on smaller scales \cite{Ly2007}, as seen in Figure \ref{fig:CWalkerM87}. If these features persist at EHT scales near M87's black hole, they will provide sharp criteria for discriminating among phenomenological models input into simulations in this 
%proposed 
work. Observing the inner regions of M87 near the base of its relativistic jet has the potential for shedding light on long-standing jet mysteries, such as the locus of its most concentrated emission or the composition-- leptonic ($e^+e^-$) or hadronic ($
%e-,
p,\mathrm{ions}$)-- of the jet plasma. 

% \subsubsection{3C 279}
The powerful quasar 3C 279 exemplifies the mystery of rapid variability in jet/accretion disk/black hole (JAB) systems. Quasar 3C 279 has a black hole with a light crossing time of an hour, yet exhibits light curves with doubling times on the scale of minutes \cite{Ackermann2016}. Variability for a wide range of phenomenological models can 
%readily 
be assessed by summing over intensity from images at different observer times with routines %described in Section 2
described in this work. Further targets of observation and modeling include 
%the distant high power radio quasar 3C 279 and 
the highly polarized FR I jet of 3C 31 (see Table 1).
\newline
$\qquad$
\newline
% %Voici comment présenter un tableau de résultats:
% Table 1. Comparison of Sgr A* distance and central black hole mass with those of other jet/accretion disk/black hole (JAB) systems
% % $$
% % % Tableau en mode math (en mode texte, il faut utiliser un {tabular})
% % \begin{array}{c|c|c} % Les barres donnent les barres verticales, les "c" pour "centrage" 
% % \text{Source}	&	\text{Distance}	&	\text{Black Hole Mass}\ 	% Les bords des cellules sont délimitées par des &
% %     \\ \hline\hline % Passage à la ligne et ligne horizontale double
% % \mathrm{Sgr A^*}
% % $\cite{Christian2015}$	
% % &	 (7.94\pm 0.42)\text{kpc}	
% % &	(4.31\pm 0.36)\times 10^6 M_\odot	
% % \\
% % \mathrm{M87} $\cite{Christian2015}$	&	(16.7\pm 0.9)\text{Mpc}	&	(6.2\pm 0.4)\times 10^9 M_\odot	\\
% % \mathrm{3C\ 31}$\cite{%Portaluri:2013
% % Woo2002}$	&	%64\text{Mpc}
% % 72\text{Mpc}	&	%5\times 10^8 M_\odot
% % 3.3\times 10^8 M_\odot	\\
% % \mathrm{3C\ 279}$\cite{Homan2000}$ $\cite{%Marscher:2004
% % Nilsson2009}$	&	1.95\text{Gpc}	&	%(4.5\pm 1.5)\times 10^8 M_\odot	
% % 7.9\times 10^8 M_\odot	 \\
% % \end{array}\label{TableOfSources}
% % $$

\begin{table}[H]
\caption{Comparison of distances to and central black hole masses of jet/accretion disk/black hole (JAB) systems}
\small % Font size can be changed to match table content. Recommend 10 pt.
\centering
\begin{tabular}{ccc}
\toprule
\textbf{Source}	& \textbf{Distance}	& \textbf{Black Hole Mass}\\
\midrule
M87 \cite{Blakeslee2009}\cite{Gebhardt2011}
%\cite{Christian2015}
&	$(16.7\pm 0.9)$\text{Mpc}	&	$(6.6\pm 0.4)\times 10^9 M_\odot$	\\
$\mathrm{3C\ 31}$ \cite{%Portaluri:2013
Woo2002}	&	%64\text{Mpc}
72
%%%
 \text{Mpc}	&	%5\times 10^8 M_\odot
$3.3\times 10^8 M_\odot$	\\
$\mathrm{3C\ 279}$ \cite{Homan2000}\cite{%Marscher:2004
Nilsson2009}	&	1.95
%%%
\text{Gpc}	&	%(4.5\pm 1.5)\times 10^8 M_\odot	
$7.9\times 10^8 M_\odot$	 \\
\bottomrule
\end{tabular}
\end{table}\label{TableOfSources}
 
%\subsection{Simulations}
Substantial progress has also been made on the computational front, as three-dimensional GRMHD simulations 
%have for nearly a decade been 
are now nearly detailed enough to resolve the magnetorotational instability 
%(MRI)
while evolving long enough 
%for JAB systems 
to exhibit 
%quasi-periodic oscillations (QPO) 
variability on many         
timescales and the stability
of
relativistic jets. The high-accuracy relativistic magnetohydrodynamics (HARM) code \cite{Gammie2003} has set the standard for evolving accretion flow dynamical variables and photon trajectories in the high-spacetime-curvature vicinity of black holes, where general relativity predicts significant distortions to the geodesic path of light.
Subsequent simulations are well adapted to the study of powerful, stable relativistic jets, e. g., 
%(McKinney and Blandford, 2009) 
\cite{McKinney2009} and 
%(McKinney, Tchekhovskoy and Blandford, 2012
\cite{McKinney2012}) (hereafter MB09 and MTB12). Other GRMHD simulations concentrate gridlines at small angles near the equatorial plane in order to analyze disk emission near the innermost orbits around the central black hole, e.g., 
%(Ressler, Tchekhovskoy, Quataert, Chandra and Gammie, 2015) 
\cite{Ressler2015} and 
%(Ressler, Tchekhovskoy, Quataert, and Gammie, 2015) 
\cite{Ressler2017}%(hereafter RTQCG15 and RTQG17)
.
%This project will focus on using these
Richard Anantua's 2016 Stanford doctoral dissertation \cite{Anantua2016} %The current work 
used a general relativistic magnetohydrodynamic (GRMHD) simulation described in MB09 and MTB12 based on the HARM code to match observational features of JAB systems at radio, optical and gamma ray wavelengths and used the same simulation to determine statistical properties of multi-wavelength emission from individual sources and surveys. This paper presents a subset of this work applied to M87 to reverse engineer observational signatures, images and polarization maps.
%%%
In our case of single-fluid GRMHD without electron thermodynamics, we neglect the effects of electron conduction on gas dynamics and assume the adiabatic index
%$\gamma_\mathrm{E.o.S.}$ 
is a function of total gas (and not electron gas) properties.
%%%
In what follows, the gravitational radius $M=GM_\mathrm{BH}/c^2$ will be used as a mass, length and time scale by setting $G=c=1$.
\section{Materials and Methods}

% This section should be divided by subheadings. Materials and Methods should be described with sufficient details to allow others to replicate and build on published results. Please note that publication of your manuscript implicates that you must make all materials, data, and protocols associated with the publication available to readers. Please disclose at the submission stage any restrictions on the availability of materials or information. New methods and protocols should be described in detail while well-established methods can be briefly described and appropriately cited.

% Research manuscripts reporting large datasets that are deposited in a publicly available database should specify where the data have been deposited and provide the relevant accession numbers. If the accession numbers have not yet been obtained at the time of submission, please state that they will be provided during review. They must be provided prior to publication.

\subsection{Simulation}
The simulation used here evolves a relativistic jet sourced by a geometrically thick disk accreting onto a rapidly-rotating ($a/M=0.92$) black hole for duration $3300M$. The data are interpolated to 256x256x256-Cartesian lattice datablocks representing physical domains ranging from 80Mx80Mx200M to 80Mx80Mx1400M to 320Mx320Mx1400M. The source code for the particular simulation "observed" can be made available on GitHub with permission of Jonathan McKinney of the University of Maryland, though the routines described are generally applicable to GRMHD jet simulations.

\subsection{Observing Simulations}

In \cite{Anantua2016}, a robust pipeline was developed to create constant observer time line-of-sight intensity and polarization maps of simulation output from arbitrary observer orientations. The MB09/MTB12 simulation used output solutions for state variables four velocity $u^\mu$, four field $b^\mu$, gas energy density $u_g$ and rest mass density $\rho$ of the GRMHD equations
%\cite{Sadowski:2014aa}
for an astrophysical plasma

\begin{equation}\label{eq:GRMHD}
\begin{tabular}{cr}
$\nabla_\mu(\rho u^\mu)=0$\\
$\nabla_\mu T_\nu^\mu=0$
\end{tabular}
\end{equation}
where $T_\nu^\mu$ is the stress tensor, 
 \begin{equation}\label{StressTensor}
 T_\nu^\mu = (\rho+u_g+p_g+b^2)u^\mu u_\nu+(p_g+(1/2)b^2)\delta^\mu_\nu - b^\mu b_\nu\end{equation}
%I have added an 
Adding equation of state 
\begin{equation}
p_e=(\gamma_\mathrm{E.o.S}-1)u_e
\end{equation}
(where $\gamma_\mathrm{E.o.S}=4/3$ for relativistic electrons) and 
%emission prescriptions to reproduce observations of JAB systems
prescriptions for emission due to a power law distribution of electrons $N_e(\gamma)\sim\gamma^{-p}$ with model-dependent relativistic electron energy density $u_e$ yields self-consistent models.

\subsection{Physical Models}
\subsubsection{Synchrotron Models}
%\subsubsection{Beta and Bias Models}
For power law synchrotron radiation, the emission scales with electron gas energy density $u_e$, magnetic field and frequency as $j_\nu\sim u_e b^{1+\alpha}\nu^{-\alpha}$, where the photon spectral index is $\alpha=\frac{p-1}{2}$ and we take $p=2$. Inspired by equipartition, a simple assumption for the electron gas energy density is that it is 
%of order 
a constant fraction of the magnetic field energy density $u_B=b^2/2$. This relation between electric and magnetic energy densities can be parameterized in a beta model where the constant $\beta\sim u_e/u_B$. A generalization of this, the "bias" model, 
%generalizes the beta model by scaling
scales $u_e$ with powers of $u_B$ as $u_e\sim \beta u_B^N$.

%\subsubsection{Current Density and Shear Jet Models}
Synchrotron radiation theory emission and absorption formulae in \cite{Anantua2016} %%%
(cf. Eq. 2.24)
%%%
conveniently express 
%%%$u_e$ in terms of the partial pressure $\tilde{P}_e$ due to electrons emitting at the observed frequency.
polarized emissivities and absorption coefficients in terms of the partial pressure $\tilde{P}_e$ due to electrons emitting in the octave around the observed frequency.
%%%
The partial pressure is then expressed as $\tilde{P}_e=Wt$%%%where 
, where $W$ is a dissipation rate per unit volume and $t$ is the characteristic (radiative or expansion) cooling time. The key advantage of casting the formulae this way is that dissipation from particularly well-motivated physical process can explicitly be added to the models. The "current density" model scales dissipation as the square of the 
%4-current density: $W\sim j_\mu j^\mu$, 
%%%
current density
\begin{equation}
W\sim j_\mu j^\mu
\end{equation}
, which simulations in MB09 and MTB12 have shown
%%%
to be
%%%
roughly the z-component of curl($\mathbf{B}$), which
%%%
is more prominent in the central jet "spine" than the peripheral enclosing "sheath." The "shear" model relates dissipation to the principle shear component of stress-energy-momentum tensor $T_{\mu\nu}$.
%%%
The dominant component of the comoving rate of velocity shear of a fluid element is
$
% \begin{equation}
S'=\left(1-\frac{v_z^2}{c^2}\right)^{-1/2}\left|\frac{dv_z}{ds}\right|\sim \left|\frac{dv_z}{ds}\right|
% \end{equation}
$, where $s$ is cylindrical radius. The shear model has 
\begin{equation}
W=\frac{1}{2}S\tau
\end{equation}
, where shear strain $\tau=\mu S$ and $\mu$ is dynamic viscosity.
%%%

\subsubsection{Inverse Compton Models}
The inverse Compton process is key for modeling gamma ray emission from the AGN selected for this investigation. Simple models in which $j_\nu\sim\mathcal{D}^4\tilde{P}_e$, where Doppler factor is the ratio of observed and fluid comoving frequencies $\mathcal{D}=\frac{\nu_\mathrm{Obs}}{\nu_\mathrm{Com}}$, yield gamma ray emission-- but gamma ray synchroton has not yet been excluded \cite{Anantua2016}. Devising models with variability timescales of 3C 279 inspired by those proposed here may help resolve this degeneracy.

\subsection{Polarized Radiative Transfer}
The process of converting emission from the aforementioned models into maps of %%%Stokes'
Stokes parameters $(I,Q,U)$ for the intensity and two independent polarizations is implemented by numerically solving the polarized radiative transfer equations in \cite{Anantua2016}. Semi-analytic integration of these radiative transfer equations has been applied to stationary, self-similar, axisymmetric models \cite{Blandford2017}, however, the 3D, time-dependent calculation by ray tracing over the simulation enables analysis of features such as apparently superluminal knots, non-axisymmetric instabilities and variability. The ray tracing code adds to the observer plane contributions to each 
%%%Stokes' 
Stokes parameter from planes at various distance/retarded time combinations along the line of sight to construct constant observer time image maps. %General relativistic effects must be added to this ray tracing scheme for MB09 and MTB12 to make comparisons with the higher frequency/smaller scale observations for the current proposal (general relativistic ray tracing has already been incorporated for simulations in RTQCG15 and RTQG17). 
The observer plane is free to rotate around its center to fix the orientation of the projected jet, and around the simulation to fix observer polar and azimuthal angles. In what follows, we stick to observer times around $2000M$ when the simulation appears transient-free and a stable jet appears Poynting-flux dominated for small radii $r\lesssim 100M$ and is gradually mass-loaded by entrained particles at larger radii.

%%%%%%%%%%%%%%%%%%%%%%%%%%%%%%%%%%%%%%%%%%
\section{Results: Comparing Jet Simulations with Observations}

% This section may be divided by subheadings. It should provide a concise and precise description of the experimental results, their interpretation as well as the experimental conclusions that can be drawn.
\subsection{Emulating Spatiotemopral Jet Properties}

\subsubsection{Collimation}
% \subsection{Comparing Simulations with Observations}
% \subsection{Matching Disk Observations}
% \subsubsection{Observational Signatures}
% %$\qquad$Located at our Galactic Center, Sagittarius A* is a prime target of observation due to its proximity. 
% Disk emission from Sgr A* has been observed and modeled throughout the electromagnetic spectrum: the X-ray bremsstrahlung emission originates from hot gas at large radii according to the advection-dominated accretion flow (ADAF) model 
% %(Narayan, Yi and Mahadevan, 1995 
% \cite{NarayanYiMahadevan1995}
% %)
% ; the inverse Compton component is near the horizon according to Moscibrodzka's models in %(Moscibrodzka et al. 2009 
% \cite{Moscibrodzka2009}
% %)
% ; and the cyclo-synchrotron sub-mm emission originates close to the black hole from models by Shcherbakov% (Shcherbakov, Penna and McKinney 2012
% \cite{Shcherbakov2012}
% %)
% . However, compared to the approach of "observing" simulations of MB09, MTB12, RTQCG15 and RTQG17, semi-analytical models
% %, e.g., (Narayan and Yi, 1995 \cite{NarayanYi1995}) 
% and other GRMHD simulations of Sgr A* to date lack the self-consistency of the unifying framework of particle acceleration, dissipation, emission, and radiative transfer.
% % that is used in the approach of observing the simulations of MB09 and MTB12.

% \subsection{Matching Jet Observations}
%\subsection{Observational Signatures}

\begin{figure}[H]\nonumber
\begin{align}
\hspace{0.0cm}  & \includegraphics[%%%height=100pt,width=140pt
height=125pt,width=170pt,trim = 0mm 1mm 0mm 0.0cm,clip]{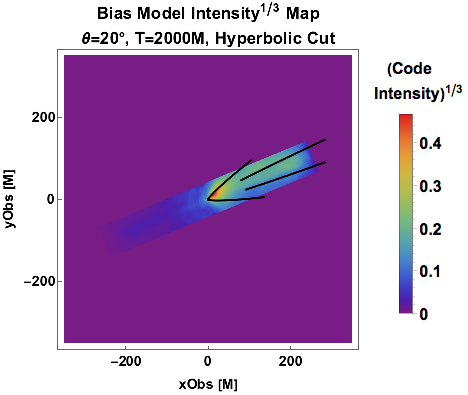%BiasModelCollimation.png
}  & 
% \end{align}  
% \begin{align}
\includegraphics[%height=100pt,width=140pt
height=125pt,width=170pt
,trim = 0mm 1mm 0mm 0.0cm,clip%height=140pt,trim = 0mm 1mm 0mm 1mm
]{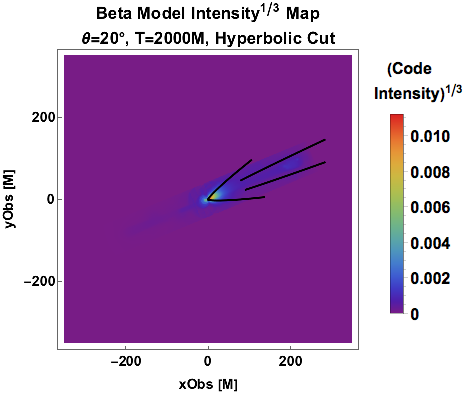%BetaModelCollimation.png
}  & 
\end{align}%\nonumber
\caption[Simulated Bias and Beta Model Intensity Maps]{
%Bias model with constant gas pressure (left panel) and beta model (right panel) viewed at 20$^\circ$ from the jet axis. The contours are the collimation profiles deduced from VLBA and EHT data. 
Intensity maps viewed at $20^\circ$ observer angle for a simulation where points away from the region $x^2+y^2-\left(\frac{z}{2M}\right)^2<(40M)^2\cup (x^2+y^2 <(40M)^2\cap |z|>40M)$ have been excised to isolate the jets. The $N=0$ bias model (left) exhibits a broader and more realistic collimation profile than the $\beta=1$ model (right) compared to the observations (thick black lines). The figures have been transformed by $(\cdot)^{1/3}$ for visual contrast.}\label{M87CollimationProfiles}
\end{figure}

Starting with the simple beta and bias models, we can develop intuition about the role the functional dependence of emissivity on magnetic field strength plays in 
%an important characteristic of jet morphology
a defining characteristic of jets:  collimation. Using the simulation datablock representing the largest physical region and assuming optical thinness, Figure \ref{M87CollimationProfiles} shows the $N=0$ bias model (where $j\sim b^{3/2}$) has a much broader jet profile than the $\beta=1$ model (where $j\sim b^{7/2}$). The contours are collimation profiles \cite{Globus2016} deduced from VLBA and EHT data. The intensity maps are left in code units as the equipartition models are not well-suited to represent the detailed emission mechanism over all jet regions displayed, though are still useful tools to illustrate changes in the way a simulation "lights up" in response to changing important dynamical variables.

%From simulation, B_phi~r^-1

\subsubsection{Jet Magnetic Field Substructure}

\begin{figure}[H]\nonumber
\begin{align}
% & \includegraphics[height=130pt,width=150pt,trim = 10mm 1mm 5mm 1mm]{xObs-yObs_bSqToThreeFourths_Intensity_Map_Rainbow_Zoom_to_IMax_ParabolicPt5_MidCut020__ThetaObs_015deg__PhiObs_000deg__Orient_202Pt5__TObs_2000M__NuObs_001.png}
& \hspace{3cm}  \includegraphics[%%%height=130pt,width=150pt
height=140pt,width=180pt,trim = 10mm 1mm 5mm 1mm]{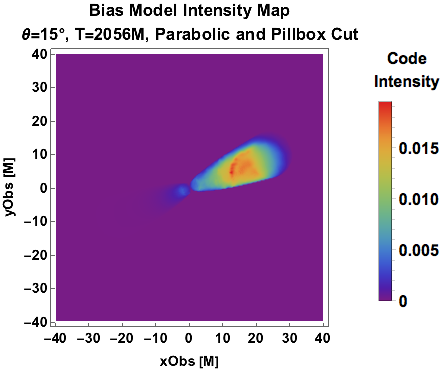} & 
%& \ \ \ \  \includegraphics[height=130pt,width=150pt,trim = 10mm 1mm 5mm 1mm]{xObs-yObs_ParabolicPtTo5_Bias_Model_Convolved__SigmaEHT_8_25__Intensity_Map_Rainbow_to_1_0IMax_ThetaObs_015deg__PhiObs_000deg__PhiOrient_202Pt5__TObs_2056M.png}
\end{align}\caption[Simulated Bias Model Intensity Map]{Jet intensity map at 15$^\circ$ viewing angle from jet axis. Emission is modeled as $j\sim (b_\mu b^\mu)^\frac{3}{4}$ ($N=0$ bias model). Using a parabolic region to subtract the disk, all regions away from the locus satisfying $0.5|z|>x^2+y^2>20$ are set to zero. A counterpropagating jet is also visible.}\label{BiasModelSimulatedSubstructure}
\end{figure}

Instabilities cause jets to pinch and kink.
%Linear instability analysis
In the observation in Figure \ref{SwirlingJet}, M87 has a distinctive swirling pattern possibly indicative of magnetic Kelvin-Helmholtz or kink instability. In Figure \ref{BiasModelSimulatedSubstructure}, viewed at 15$^\circ$ using a parabolic geometric jet isolation, we see a similar corkscrew feature for the $N=0$ bias model for the highest resolution simulation lattice. Again, the normalization of the intensity map is irrelevant, as here we are concerned with intensity contrast, which illustrates the magnetic field substructure in the jet. 

Also noteworthy is that there is a relatively dim counterjet in Figure \ref{BiasModelSimulatedSubstructure}, which, as in the observation Figure \ref{fig:CWalkerM87}, indicates preferential Doppler beaming towards the observer direction.

\subsubsection{Polarization Maps}

\begin{figure}[H]\nonumber
\begin{align}
%  & \includegraphics[height=130pt,width=150pt,trim = 10mm 1mm 5mm 1mm]{xObs-yObs_ParabolicPtTo5_Bias_Model_Intensity_Map_Rainbow_to_1_0IMax_ThetaObs_015deg__PhiObs_000deg__PhiOrient_202Pt5__TObs_2056M.png} 
&  \includegraphics[%%%height=130pt,width=150pt
height=145pt,width=180pt,trim = 10mm 1mm 5mm 1mm]{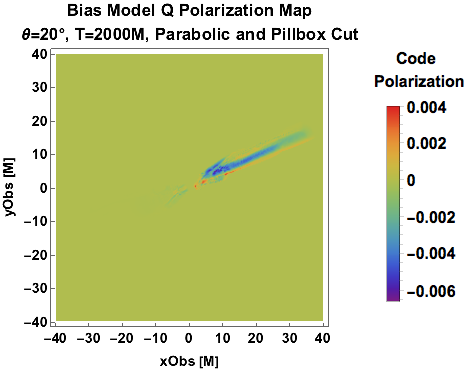} & & \ \ \ \ 
  \includegraphics[%%%height=130pt,width=150pt
  height=145pt,width=180pt,trim = 10mm 1mm 5mm 1mm]{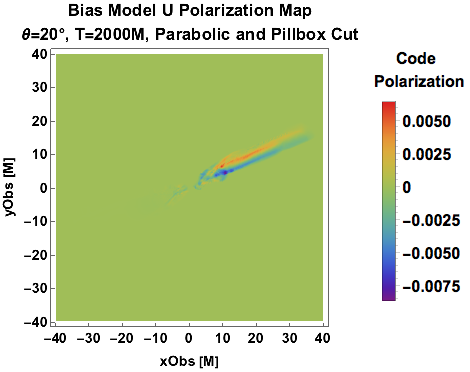}
%   & & 
% \includegraphics[height=80pt,trim = 10mm 1mm 0mm 1mm]{UbSqTo1Pt5_Intensity_Map_to_MaxVal_Rainbow_Zoom_ThetaObs_015deg__PhiObs_000deg__tObs_2000M_png__NuObs_001.png}
\end{align}\caption{Linear polarizations $Q$ (left) and $U$ (right) maps at 20$^\circ$ viewing angle generated from the 
%same 
emissivity function $j\sim (b_\mu b^\mu)^\frac{3}{4}$
and parabolic geometric jet region isolation.}\label{ParabolicJetPolarization}
\end{figure}

Due to the polarized nature of synchrotron radiation, it is essential to include polarization in our "observations" of jet emission for synchrotron models. In addition to total intensity, the "Observing" Jet Simulations pipeline can compute maps of other %Stokes’ parameters
Stokes parameters as in Figure \ref{ParabolicJetPolarization}, where we have Stokes maps of two linear polarizations $Q$ (left) and $U$ (right) for an optically thin $N=0$ bias model at the highest simulation resolution. The $U$ map is limb brightened and has a region of low polarization near the spine as $U$ changes sign. The $Q$ map is orthogonal to $U$.

% \begin{figure}[H]
% \begin{center}
% % Inclusion d'une image: ce qui est entre crochets est un argument "optionnel", alors que l'argument obligatoire est entre accolades
% \includegraphics[height=150pt,width=200pt]{xObs-yObs_ParabolicPtTo5_Bias_Model_Intensity_Map_Rainbow_to_1_0IMax_ThetaObs_015deg__PhiObs_000deg__PhiOrient_202Pt5__TObs_2056M.png} 
% \caption{Jet image map at 15$^\circ$ viewing angle from jet axis. Emission is modeled as $j\sim (b_\mu b^\mu)^\frac{3}{4}$. Using a parabolic region to subtract the disk, all regions away from the locus satisfying $|z|>x^2+y^2>20$ are set to zero. Counterpropagating jet is also visible.}\label{ParabolicJet}
% \end{center}
% \end{figure}
 
\subsubsection{Instrument-Specific Properties: Cadence and Convolution}

\begin{figure}[H]\nonumber
\begin{align}
 & \includegraphics[%%%height=110pt,width=130pt
 height=130pt,width=135pt,trim = 10mm 1mm 5mm 1mm]{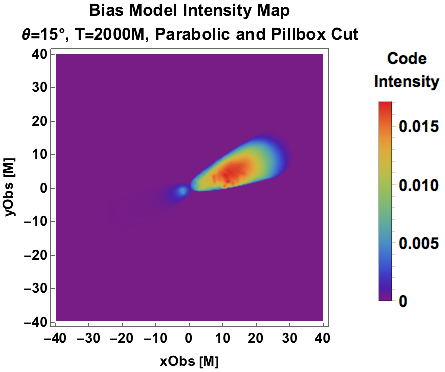} & \ \ \includegraphics[%%%height=110pt,width=130pt
 height=130pt,width=135pt,trim = 10mm 1mm 5mm 1mm]{xObs-yObs_ParabolicPtTo5_Bias_Model_Intensity_Map_Rainbow_to_1_0IMax_ThetaObs_015deg__PhiObs_000deg__PhiOrient_202Pt5__TObs_2056M.png} & & \ \  
  \includegraphics[%%%height=110pt,width=130pt
  height=130pt,width=135pt,trim = 10mm 1mm 5mm 1mm]{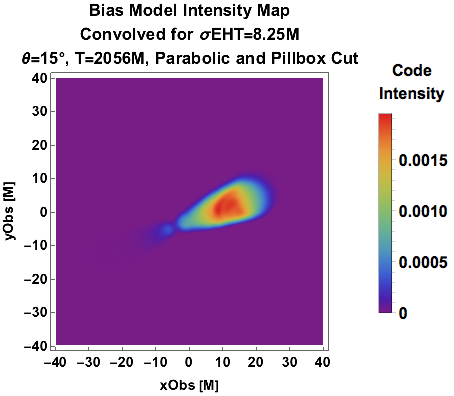}
%   & & 
% \includegraphics[height=80pt,trim = 10mm 1mm 0mm 1mm]{UbSqTo1Pt5_Intensity_Map_to_MaxVal_Rainbow_Zoom_ThetaObs_015deg__PhiObs_000deg__tObs_2000M_png__NuObs_001.png}
\end{align}\caption[Simulated Bias Model Intensity Maps]{Jet intensity maps viewed at $15^\circ$ observer angle for a simulation where points away from the region $0.5|z|>x^2+y^2>20$
%$x^2+y^2>|z|>20M$ 
have been excised to isolate the jet and counterjet for the $N=0$ bias model. The left map is taken at observed time $T=2000M$, the middle map shows significantly variation in the form of increasingly ordered substructure at $T=2056M$ and the right map displays how the structure is degraded by convolution with a $\sigma=8.25M$ Gaussian beam.}\label{BiasModelSimulated}
\end{figure}

The process of observing itself fundamentally affects the astrophysical inferences that can be drawn from the dynamical evolution of a source and its surroundings. The convolution and cadence of images due to an observing instrument set minimum spatial and temporal scales that can be probed. For concreteness, the 
%%%VLA
VLBA 43 GHZ observations \cite{Ly2007} occurred with a cadence of an image per 21 days, which for M87 corresponds to $56M$ (56 light crossing times of a gravitational radius). In the spatial domain, the point spread function for the EHT is $8.25\ \mu$as. Here it is noteworthy that there is significant overlap between the spatiotemporal scales of observations and simulations, as illustrated with the bias model "observed" at different times and with/without convolution in Figure
\ref{BiasModelSimulated} (for the highest simulation resolution). The jet substructure clearly varies on the timescale of tens of $M$ and/or with convolution by a Gaussian beam with width of order unity.  

\subsection{Matching Models with Observed Images}
 
\begin{figure}[H]\nonumber
\begin{align}
& 
\includegraphics[height=100pt,trim = 6mm 1mm 0mm 1mm]{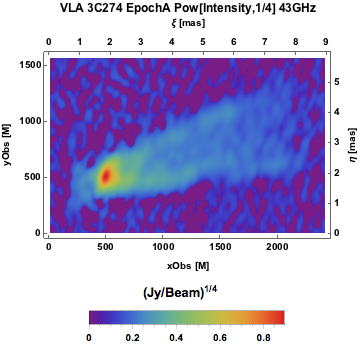} & 
\includegraphics[height=100pt,trim = 6mm 1mm 0mm 1mm]{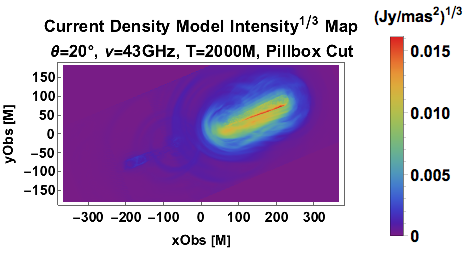} & 
\includegraphics[height=100pt,trim = 6mm 1mm 0mm 1mm]{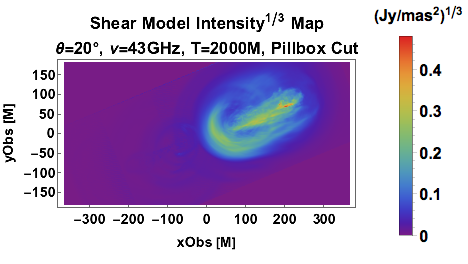} &
\end{align}
\caption{M87 VLA observation (left)
%movie snapshot at $t_\mathrm{Obs0}$  (left column) side-by-side with a snapshot at $t_\mathrm{Obs0}+10t_\mathrm{Step}$ (right column) for  
%$t_\mathrm{Step} = 21\mathrm{days}\approx 56M_\mathrm{M87}$
at 43 GHz. The total intensity is 
%shown on the top row, and intensity is 
monotonically transformed by $(\cdot)^{1/4}$ for visual clarity %on the bottom row
. The beam dimensions are $4.3\cdot10^{-4}\mathrm{arcs}\cdot 2.1\cdot10^{-4}\mathrm{arcs}$. 
%These images can be viewed as a movie sans transformation, courtesy of Craig Walker and his collaborators, here:
% \url{http://www.aoc.nrao.edu/~cwalker/M87/}.
Synthetic 43GHz observations are shown for comparison using the current density (middle) and shear (right) models and transformed by $(\cdot)^{1/3}$.
}
\label{Jets43GHzComparison}
\end{figure}

Having made some suggestive comparisons of discrete observational signatures with optically thin toy models, we have the necessary background to use the full polarized radiative transfer routines with opacity to replicate observations of a particular source, M87, at a particular frequency, attempting to match all measured properties including: morphology, orientation, and normalized intensity.  To this end, we employ the current density and shear models in a comparison to the 43 GHz %%%VLA  
VLBA observation in Figure \ref{Jets43GHzComparison}. One limitation is that the observed image spans a greater spatial extent than the synthetic image, though this can be remedied in the near future when detailed observations are produced on linear scales a factor of order unity smaller and/or detailed simulations extend on scales a factor of order unity greater. In the current density model, we see a bright spine at small cylindrical radius, and we also see current layers in a corona at the jet boundary. The shear model prescription, in which the partial pressure scales as the square of velocity shear, appears relatively more edge brightened, suggesting shear is prominent at the boundary.

% The more realistic current density and shear models have greater morphological resemblance to the observations, as seen the the comparison at 43 GHz in Figure \ref{Jets43GHzComparison}. 

\section{Discussion}

% This section may be divided by subheadings. Authors should discuss the results and how they can be interpreted in perspective of previous studies and of the working hypotheses. The findings and their implications should be discussed in the broadest context possible. Future research directions may also be highlighted.

The intensity map in Figure 
%\ref{ParabolicJet}
\ref{BiasModelSimulatedSubstructure} generated from the highly stylized bias model-- in which relativistic gas pressure is constant along jets-- already shows the helical jet substructure and counterjet visible 
%\ref{SimulationTObs2000MSnapshotsbSqToThreeFourthsUgOnRhoGtr1Em2And5Em2And1E0}
in the observation in Figs. \ref{SwirlingJet} and \ref{fig:CWalkerM87}, respectively. The more realistic current density and shear models have greater morphological resemblance to the observations, as seen though the comparison at 43 GHz in Figure \ref{Jets43GHzComparison}. Though the models significantly vary from the observations at this stage, we are now better equipped to interpret why different regions of the jet light up in different ways. For example, the current density jet is brightest where in the simulation the current density-- roughly the z-component of the curl of the magnetic field, is greatest; the shear model can be interpreted as accelerating particles most in regions of high velocity shear. The ultimate test of the "Observing" Jet Simulations methodology is whether the right combination of model features can generate synthetic observations nearly indistinguishable from real observations.

Armed with a robust "Observing" Jet Simulations pipeline from GRMHD simulations to synthetic observations \cite{Anantua2016}, the next target is the post-processing of a broad range of astrophysical simulation data matching new observations of JAB systems (with possible extensions to proto-planetary disks, white-dwarf binaries, etc). One of the nearest jets is in M87, which has been observed in the gamma ray band by Fermi Large Area Telescope, in the optical band by the Hubble Space Telescope and the radio band by the ALMA. M87 is an excellent target of this investigation because of the wealth of existing and planned mm- and sub-mm very long baseline (VLB) EHT observations, and the fact that its high central black hole mass makes it the only AGN with jets and a central black hole of comparable angular width from Earth as Sgr A*. 
Applying this methodology to other sources such as 3C 279 and 3C 31 may elucidate the most likely mechanisms underlying phenomena such as rapid variability and large-scale polarization reversals.

%%%%%%%%%%%%%%%%%%%%%%%%%%%%%%%%%%%%%%%%%%
\section{Conclusions}

% This section is not mandatory, but can be added to the manuscript if the discussion is unusually long or complex.

The "Observing" Jet Simulations methodology enables us to self-consistently simulate multiwavelength astronomical observations including a wide range of effects including Doppler beaming, polarization, opacity, and to rotate the observer to simulate many sources. The approach used here emphasizes models with clear physical interpretations such as equipartition, current density and shear. In addition to more physical models, we expect greater dynamical range in simulations and observations  to come.

%%%%%%%%%%%%%%%%%%%%%%%%%%%%%%%%%%%%%%%%%%
\vspace{6pt} 

% %%%%%%%%%%%%%%%%%%%%%%%%%%%%%%%%%%%%%%%%%%
% %% optional
% \supplementary{The following are available online at www.mdpi.com/link, Figure S1: title, Table S1: title, Video S1: title.}

%%%%%%%%%%%%%%%%%%%%%%%%%%%%%%%%%%%%%%%%%%
\acknowledgments{Richard Anantua is supported by the California Alliance. Jonathan McKinney has provided the simulation used in this work.
%All sources of funding of the study should be disclosed. Please clearly indicate grants that you have received in support of your research work. Clearly state if you received funds for covering the costs to publish in open access.
}

%%%%%%%%%%%%%%%%%%%%%%%%%%%%%%%%%%%%%%%%%%
\authorcontributions{Richard Anantua has carried out the calculations presented in this work and has written this manuscript. Alexander Tchekhovskoy has guided Richard Anantua through the initial stages of radiative transfer code development.  Roger Blandford has has conceived, designed, managed and coordinated this project.
%For research articles with several authors, a short paragraph specifying their individual contributions must be provided. The following statements should be used ``X.X. and Y.Y. conceived and designed the experiments; X.X. performed the experiments; X.X. and Y.Y. analyzed the data; W.W. contributed reagents/materials/analysis tools; Y.Y. wrote the paper.'' Authorship must be limited to those who have contributed substantially to the work reported.
}

%%%%%%%%%%%%%%%%%%%%%%%%%%%%%%%%%%%%%%%%%%
\conflictofinterests{The authors declare no conflict of interest.
%Declare conflicts of interest or state ``The authors declare no conflict of interest.'' Authors must identify and declare any personal circumstances or interest that may be perceived as inappropriately influencing the representation or interpretation of reported research results. Any role of the funding sponsors in the design of the study; in the collection, analyses or interpretation of data; in the writing of the manuscript, or in the decision to publish the results must be declared in this section. If there is no role, please state ``The founding sponsors had no role in the design of the study; in the collection, analyses, or interpretation of data; in the writing of the manuscript, and in the decision to publish the results''.
}

%%%%%%%%%%%%%%%%%%%%%%%%%%%%%%%%%%%%%%%%%%
%% optional
\abbreviations{The following abbreviations are used in this manuscript:\\

\noindent 
AGN: Active Galactic Nucleus\\
EHT: Event Horizon Telescope\\
GRMHD: General Relativistic Magnetohydrodynamic\\
JAB: Jet/Accretion Disk/Black Hole\\
MDPI: Multidisciplinary Digital Publishing Institute\\
NRAO: National Radio Astronomy Observatory\\
OVRO: Owens Valley Radio Observatory\\
% VLA: Very Large Array\\
VLBA: Very Long Baseline Array\\
VLBI: Very Long Baseline Interferometry\\
% DOAJ: Directory of open access journals\\
% TLA: Three letter acronym\\
% LD: linear dichroism
}

% %%%%%%%%%%%%%%%%%%%%%%%%%%%%%%%%%%%%%%%%%%
% %% optional
% \appendix
% \section{}
% The appendix is an optional section that can contain details and data supplemental to the main text. For example, explanations of experimental details that would disrupt the flow of the main text, but nonetheless remain crucial to understanding and reproducing the research shown; figures of replicates for experiments of which representative data is shown in the main text can be added here if brief, or as Supplementary data. Mathemtaical proofs of results not central to the paper can be added as an appendix.

% \section{}
% All appendix sections must be cited in the main text. In the appendixes, Figures, Tables, etc. should be labeled starting with `A', e.g., Figure A1, Figure A2, etc. 

% %%%%%%%%%%%%%%%%%%%%%%%%%%%%%%%%%%%%%%%%%%
\bibliographystyle{mdpi}

%=====================================
% References, variant A: internal bibliography
%=====================================
\renewcommand\bibname{References}

%=====================================
% References, variant B: external bibliography
%=====================================
%\bibliography{your_external_BibTeX_file}

% %%%%%%%%%%%%%%%%%%%%%%%%%%%%%%%%%%%%%%%%%%
% %% optional
% \sampleavailability{Samples of the compounds ...... are available from the authors.}

%%%%%%%%%%%%%%%%%%%%%%%%%%%%%%%%%%%%%%%%%%

\end{document}